# The Necessity of International Particle Physics Opportunities for American Education

Enrique Arce-Larreta

*West High School, Salt Lake City, UT 84103*

Ketevi Assamagan

*Brookhaven National Laboratory, Upton, NY 11973*

Emanuela Barzi

*Fermi National Accelerator Laboratory, Batavia, IL 60510, and*

*The Ohio State University, Columbus, OH 43210*

Uta Bilow

*Technicshe Universität Dresden, 01069 Dresden, Germany*

Kenneth Cecire

*University of Notre Dame, Notre Dame, IN 46556*

Sijbrand de Jong

*Radboud Universiteit, 6525 AJ Nijmegen, Netherlands*

Simone Donati

*University of Pisa, Pisa 56127, Italy*

Steven Goldfarb

*University of Melbourne, Parkville VIC 3010, Australia*

Joel Klammer

*Concordia International School Shanghai, Shanghai, China*

Azwinndini Muronga

*Nelson Mandela University, Gqeberha, 6019, South Africa*

Maria Niland

*Patuxent High School, Lusby, MD 20657*

ABSTRACT

This Snowmass2021 Contributed Paper addresses the role of the Particle Physics community in creating and fostering international connections in American education. It describes the pressing need to introduce students and faculty to the challenges and rewards of international collaboration, not only to develop the next generation of scientists and engineers for particle physics, but to maintain and build U.S. leadership on an increasingly competitive world stage.

We present and assess current efforts in education and public engagement with an eye toward identifying those activities in need of change or increased resources to improve audience reach and program efficacy. We also consider possible new activities that might improve upon or complement existing programs, with the common goal of providing all U.S. students with the opportunity to benefit from a quality international scientific experience.

## 1. EXECUTIVE SUMMARY AND RECOMMENDATIONS

### 1.1. Summary

Particle physics is an international endeavor. This is driven in part by the universal nature of the scientific goals and achievements, but also by the growing experimental challenges of the field. No one institution or nation can assemble the resources or expertise needed to explore the frontiers of the field without help from the international community. Nor would it want to. Scientists in particle physics understand that the global nature of the field is one of its greatest strengths.

The diversity of national, social and cultural backgrounds present in the experiments and labs enriches the pool of intellectual thought and solidifies the validity of their scientific findings.

This fact is one of the primary messages scientists share when mentoring young students. It is not only enlightening, but it is fundamental to the scientific process. Unfortunately, for reasons of physical distance, economics and cultural practice, many students in the U.S. do not get exposure to such experiences, which are much more prevalent overseas. This leaves them at an important disadvantage, not only limiting their career choices, but also their ability to thrive in a scientific and economic environment that is becoming increasingly global every day.

Our recommendations focus on creating and enhancing opportunities for American students to receive positive and enduring exposure to the world of international scientific research. Whether or not these students continue on a path to become scientists, such experiences will be life-changing, providing them with a more comprehensive worldview and giving them the tools they will need to compete and thrive on our interconnected planet.

The involvement by U.S particle physicists based in national laboratories and at institutions of higher learning brings richness to the culture and education in America. It is also an opportunity for our particle physicists to contribute meaningfully to underdeveloped regions of the world. Participating in physics education in these regions is important for the future of our field globally and it will in turn strengthen our position as global leaders.

### 1.2. Findings and Recommendations

In the main body of this paper, we make several recommendations based on what we have found from study of the topic. The findings include the deep connections of U.S outreach and education to similar efforts in other countries and the centrality of these connections to the work, much as is true in particle physics research. Another set of important findings is that programs which enable high school students, teachers, undergraduates, or graduate students to work and learn alongside their counterparts from other countries are highly effective in terms of both the science and the ability to work in collaboration. We thus recommend strong support for and expansion of these programs as well as new steps to offer U.S.-based international education and research experiences.

## 2. INTRODUCTION

Over the past century, parallel to the birth and development of modern physics, our world has become a closely interconnected neighborhood. Revolutions in science, agriculture, transportation and communication have shrunk our planet to a point that we are in instant contact with nearly anyone on any continent. Food, energy and other vital supplies are produced and shared on a global scale, taking advantage of regional supplies in expertise, commodities and human resources to keep up with demand in an increasingly effective manner.

With these changes, it has become increasingly important that new generations of students be exposed to scientific activities that can prepare them to grow and thrive in an international environment. This will be important for their success as contributing citizens, regardless of their eventual choice of career. For the U.S. to remain a world leader in economic and scientific advancement, it will need a well-trained workforce of citizens capable of collaborating and competing on a global level. [1]

Researchers in particle physics have been well-positioned to provide such training for many decades now. The scale and resource demands of their research facilities are well beyond the reach of any single institute or even national effort.

Furthermore, over the years, the physicists have discovered and benefited from the advantages that come with international collaboration. Working with scientists and engineers from a variety of national, social and cultural backgrounds not only increases the resources available, but brings new ideas, values and methodologies that strengthen the research program significantly. [2]

Programs that bring students, both from the U.S. and abroad, in contact with this environment teach them both the universality of the scientific method and the rich diversity of approaches that can be used to understand and solve our most complex problems. They not only learn that all humanity is driven by a common quest to understand the universe, but that scientific measurements transcend human biases and opinions. Furthermore, students attending programs in an international setting gain confidence working alongside their peers and are more likely to become leaders in future global endeavors.

In the sections below, we will present a modest survey of efforts organized by researchers in experiments and laboratories, national and international coordinating bodies, international secondary schools in the U.S. and abroad, summer and semester exchange programs for undergraduates, graduate student and postdoctoral exchange programs, and secondary school teacher training programs. In all cases, we evaluate the existing programs and suggest improvements and/or complementary programs that could improve quality and overall reach, in terms of total numbers, geography and diversity of the participants.

## 3. NETWORKS AND ORGANIZATIONS

### 3.1. QuarkNet

QuarkNet, which began in 1999, continues to be the premier education program for broader impacts in particle physics in the United States. The focus is teacher professional development: each teacher affects many students and is able to participate in the program on a long-term basis, thus enabling lasting positive change. In this ongoing work, the QuarkNet staff and mentors collaborate to provide significant development opportunities for the teachers; they, in turn, collaborate with mentors and staff to provide learning opportunities for their students. [3]

Since particle physics is by its nature international, so too are many programs and opportunities in QuarkNet. QuarkNet actively collaborates with particle physics education and outreach programs in other countries and makes its resources available worldwide.

The most prominent international opportunity for QuarkNet teachers is in CERN Summer Programs, in particular the High School Teacher (HST) and International Teacher Weeks (ITW) programs [4], described elsewhere. Since the last Community Summer Study in 2013, QuarkNet, working with the University of Michigan RET program, has sent 31 teachers to CERN in this way. These teachers not only learned a great deal about particle physics but also collaborated with their colleagues from around the world, sometimes continuing that collaboration over a span of years.

Less prominent but also important is funding from the University of Notre Dame for collaborative exchanges with partner institutions around the world. Notre Dame QuarkNet teachers have gained invaluable experience traveling with staff, students, and faculty to assist in education and outreach programs in Chile, Mexico, Hong Kong, Macao, Taiwan, and Japan [5].

Perhaps the most fruitful areas of international collaboration are those created for students in collaboration with partners in various countries. In QuarkNet, teachers are asked to participate in preparing and facilitating these programs; thus they not only provide international experiences in particle physics for students but significant and ongoing international professional development for teachers. This is particularly true for International Masterclasses and Cosmic Ray Studies.

There is a "QuarkNet model" of engagement in International Masterclasses, which begins with in-class activities given by teachers to prepare students for the masterclass. QuarkNet recommends three hours of engagement in specific activities designed to help students understand how particle physics is done and, along the way, learn a few key concepts, such as the idea of an organized scheme of fundamental particles, the importance of indirect measurement, and the use of conservation laws. These give the students a greater familiarity with what they encounter in the masterclass; research indicates that it enhances the learning experience of the students. The teachers are also involved in the planning of the masterclass and in helping students as they analyze masterclass data. While the Quarknet mentor is the masterclass expert, teachers can be part of the team to increase their lever arm. This sort of engagement by teachers is itself a form of professional development. It is unique to QuarkNet but is also spread by QuarkNet internationally. [6, 7]

Students and teachers participate in International Masterclass videoconferences where they work with physicists and other students who have analyzed the same or similar data, providing direct international engagement.

Similarly, teachers with cosmic ray detectors or using the Cosmic Ray e-Lab engage with students in understanding what cosmic rays are, how detectors work, and in the design of research projects. The QuarkNet cosmic ray detector and e-Lab are used worldwide and thus, like masterclasses, provide international engagement for students. [8]

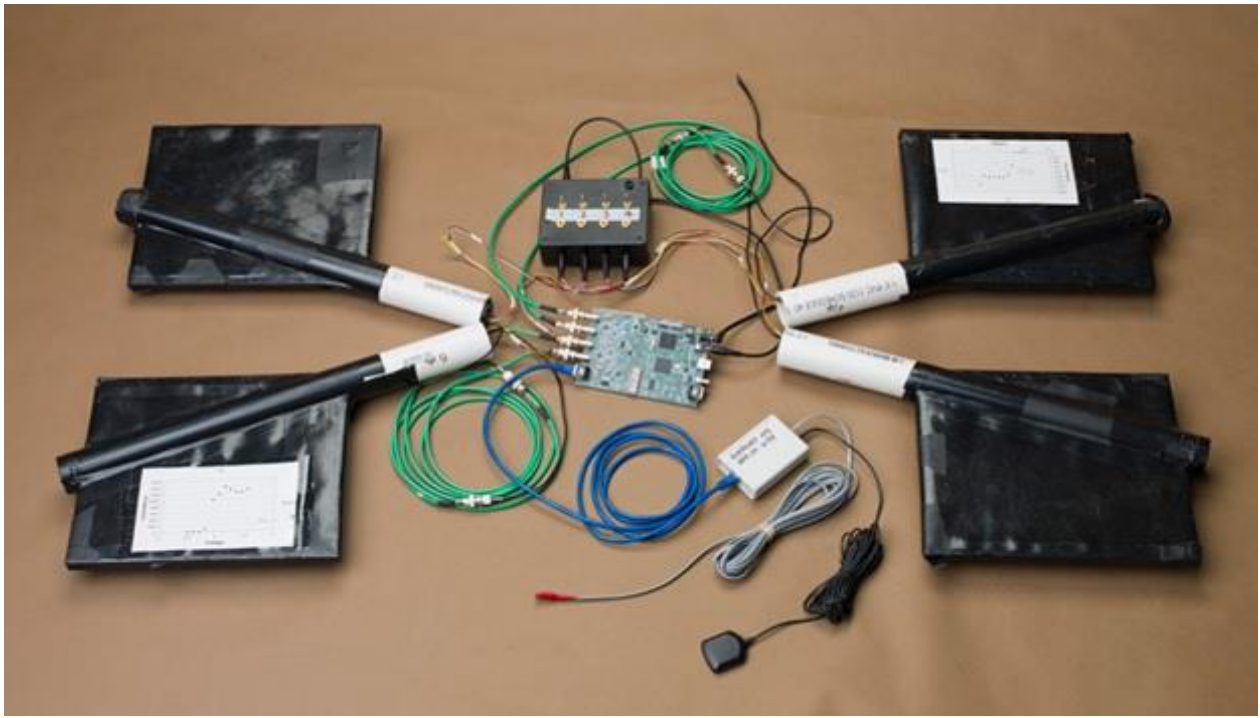

Figure 1: QuarkNet cosmic ray detector, with four scintillating counters, DAQ, GPS, voltage box, and cables.

QuarkNet also sponsors World Wide Data Day, held each autumn, which enables teachers to run a simple LHC measurement in their classrooms with their students and then connect with other students and mentors from around the world.

QuarkNet is engaged internationally with the International Particle Physics Outreach Group (IPPOG) and International Masterclasses as well as with many sister programs, most prominently Netzwerk Teilchenwelt in Germany. In this way, QuarkNet shares but also gains expertise and ideas; just as important, QuarkNet is a good collaborator in worldwide particle physics education and outreach. The U.S. representative to IPPOG is the QuarkNet staff member from Fermilab; previously this was the QuarkNet PI from Fermilab, who was also an IPPOG co-Chair. Another QuarkNet staff member is co-Coordinator of IMC along with his colleague from Netzwerk Teilchenwelt and Technische Universität Dresden. These two are both regular IPPOG contributors. Another QuarkNet staff member works with Global Cosmics in IPPOG. QuarkNet has contributed the CMS Masterclass, World Wide Data Day, International Muon Week, and cosmic ray detectors to international education and outreach. It has assisted and learned from not only Netzwerk Teilchenwelt but also groups as diverse as QuarkNet-TW in Taiwan, Tan-Q in Japan, and QuarkNet-Cymru in the UK. Most recently, QuarkNet staff and teachers have been working with the African School of Fundamental Physics and Applications and with the new Life Lab Foundation QuarkNet-India.

**Finding:**

F1. QuarkNet is at the center of international pre-university particle physics international collaboration.

**Recommendations:**

R1. QuarkNet and other U.S. outreach programs in particle physics should expand ties to international partners, particularly in the developing world.

R2. International education and outreach efforts such as International Masterclasses and various cosmic ray projects should expand and U.S. institutions should support and participate in them at increasing levels.

### 3.2. IPPOG

The International Particle Physics Outreach Group (IPPOG) is a collaboration of scientists, educators and communication specialists supporting worldwide science education and public engagement for particle physics. Its members currently include 32 countries, 6 experiments and one international laboratory (CERN). Two national laboratories, DESY and GSI, participate as associate members. [9. 10]

Although IPPOG originated as a European entity, formed in 1997 under the auspices of the European Committee for Future Accelerators (ECFA) and the High Energy Particle Physics Board of the European Physical Society (EPS-HEPP), its membership now spans the globe. Non-European members include Brazil, Israel, South Africa and Australia, as well as the U.S. QuarkNet is the signing body of the U.S., which pays an annual fee of 5000 EUR and is represented by Spencer Pasero of the University of Notre Dame.

IPPOG and QuarkNet currently have a very productive and complementary relationship. Members of their executive or advisory boards have served on or are currently serving on each other's equivalent bodies, and their major global programs are developed, organized, coordinated and communicated together. This includes the International Particle Physics Masterclasses, International Cosmic Day, International Muon Week and World-Wide Data Day. Ken Cecire, serves as coordinator of International Masterclasses for both QuarkNet, and alongside Uta Bilow, for IPPOG.

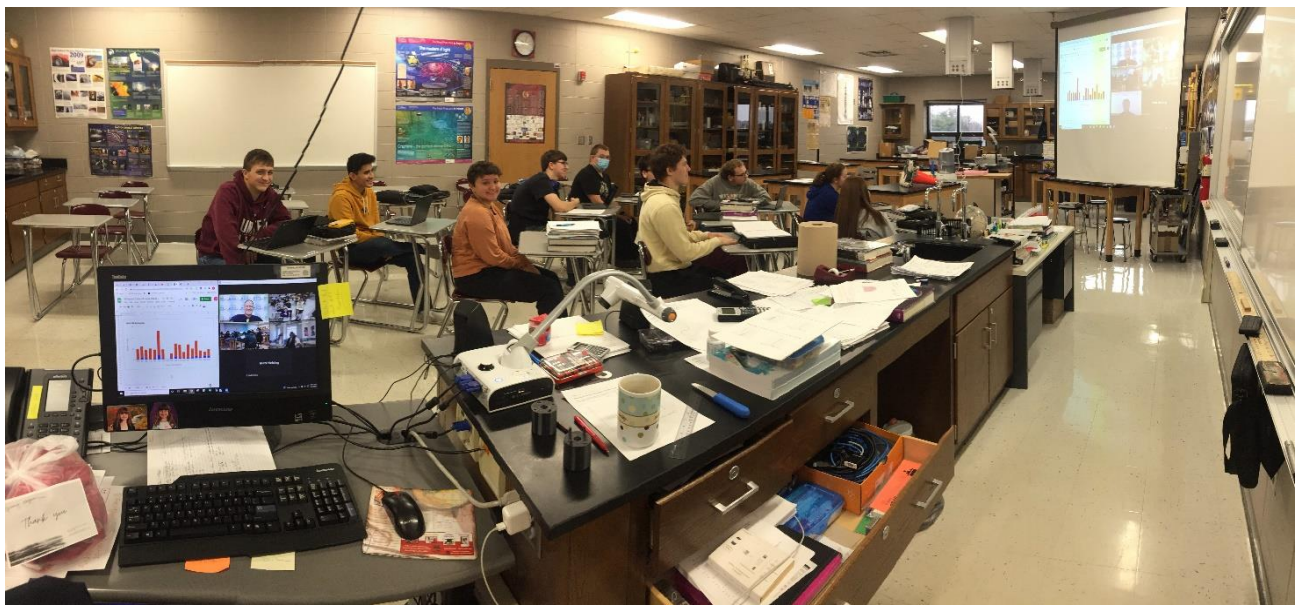

Figure 2: Students at Winamac Community HS in Indiana participate in a World Wide Data Day videoconference.

As with QuarkNet, the primary goal of IPPOG is to facilitate the ongoing work of the many active researchers around the world already engaged in particle physics education and public engagement. The representatives meet twice yearly to share new experiences and best practices, and to help to develop, improve and expand the reach of the global programs. IPPOG was acknowledged in the two most recent European Particle Physics Strategy Updates as critical to global education and public engagement, warranting support from the international HEP community.

The recent addition of national laboratories as IPPOG associate members provides a means for augmenting participation of countries heavily invested in HEP research. These laboratories commit modest monetary or human resources and assign representatives to participate in discussion on the Collaboration Board. Although only the national representative is allowed to vote, such participation does well to direct interest for IPPOG programs and commitments.

IPPOG has recently hosted several presentations and a panel discussion on its commitment to formal education and support for this was a recommendation of the 2020 European Strategy document. Furthermore, a number of national or state public education departments around the world are looking into the usage of IPPOG Masterclasses in secondary classrooms.

**Finding**:

F2. Associate membership in IPPOG by large national laboratories in the U.S. is a great benefit to particle physics education. In addition to Fermilab, other national laboratories should seek IPPOG associate membership.

### 3.3. European Networks

There are various funding opportunities through the Horizon 2020 Research and Innovation Framework Program of the European Commission (Horizon Europe since 2021). These are typically trilateral EU-US-Japan collaborations. Two examples are the 4-year long Marie Skłodowska-Curie Research and Innovation Staff Exchange (RISE) and the Marie Skłodowska-Curie Innovation Training Network (ITN). [11] ITN supports up to 15 new researchers at European

Institutions for three years each. RISE covers the travel costs of existing European researchers. In both cases, European researchers can spend up to about a third of their time in U.S. partner institutions.

Examples of successful RISE grants include:

- Muon campus in the U.S. and European contributions (MUSE, 2016) for design and construction of detector systems for Mu2e and Muon g-2 with researchers from Italy, Germany, Greece and the UK at Fermilab
- NEw WindowS on the universe and technological advancements from trilateral EU-US-Japan collaboration (NEWS, 2018) for technical aspects of Muon g-2 and Mu2e, as well as advanced superconducting technologies for particle accelerators and detectors at Fermilab, SLAC, Caltech, the NASA Marshall Space Flight Center
- INTENSE (2019) for work on neutrino detectors and physics, and Mu2e detectors, with a societal extension of muon beam applications to geophysics and archeology with researchers from Italy, UK, France, Switzerland, Belgium, Israel, Japan and US at Fermilab [12]
- PROBES (2021) proposal for Lepton Flavor Violation experiments in the neutral sector and neutrino oscillations within the Short Baseline Neutrino (SBN) program, Charged Lepton Flavor Violation experiments at Mu2e, and new accelerator technologies for high intensity and brightness muon particle beams in collaboration with PSI, based at Fermilab.

In FY20 an ITN-type network was approved and funded on Mu2e, neutrinos and neutrino detectors with 8 newly hired European researchers. The eight researchers can spend up to one year at FNAL within the four-year duration of the grant.

Not only do these endeavors require cutting-edge technologies that will hopefully open new windows in physics and technology, but they also offer an ideal way for nations to collaborate on science.

### 3.4. Additional Information on Partners in International Physics Education

Some of the important partners to U.S. international education and outreach in particle physics include:

- *American Association of Physics Teachers*. AAPT has a Committee on International Physics Education and a Contemporary Physics Committee. In addition, AAPT meetings are an important forum for dissemination of international physics education and outreach efforts in the U.S. [13]
- *American Physical Society*. APS has a strong influence through the Forum on Education, the Forum on International Physics, and the Division of Particles and Fields. [14]
- *European Physical Society*. The influence of EPS is analogous to that of APS. [15]
- *Life Lab Foundation QuarkNet-India*. Life Lab Foundation is a physics education program in India started by a CERN physicist from India. It has partnered with QuarkNet to create a program for Indian high school physics teachers to interact online with physicists, QuarkNet staff, and U.S. physics teachers. [16]
- *Netzwerk Teilchenwelt*. Netzwerk Teilchenwelt (Particle World Network) is the analogue in Germany to QuarkNet; the two programs collaborate in various ways but especially on International Masterclasses and Global Cosmics. The Co-Coordinators of IMC are staff members of Netzwerk Teilchenwelt and QuarkNet; International Cosmic Day is run though DESY and Netzwerk Teichenwelk while International Muon Week is from Fermilab and QuarkNet. [17]

- *Perimeter Institute*. Perimeter Institute in Canada has online and in-person workshops that are open to teachers from many countries and publishes physics education activities. Both are popular with U.S. teachers. [18]
- *Physics without Frontiers*. This program of ICTP brings masterclasses to developing areas worldwide, often for undergraduate students. IPPOG and, therefore, U.S. programs offer the same support given to masterclasses in general and work with them on strategies to expand the reach of IMC. [19]
- *QuarkNet-TW*. QuarkNet-TW is a cosmic ray program using QuarkNet cosmic ray detectors for students and teachers in Taiwan. They also do outreach to Macao and collaborate with QuarkNet. [20]
- *TanQ and Accel Kitchen*. TanQ (Exploration Q in English) is the cosmic ray program in Japan, which uses the QuarkNet cosmic ray detector, the Osechi ("lunchbox") detector developed at KEK, and the TanQ detector based on the Cosmic Watch developed at MIT. It is part of a broader effort, Accel Kitchen, to bring contemporary physics to Japanese high school students and teachers. TanQ and QuarkNet currently collaborate on cosmic ray studies and international masterclasses. [271]

## 4. INTERNATIONAL OUTREACH AND EDUCATION PROGRAMS

The International Outreach and Education programs in which U.S. students and educators participate most are International Masterclasses (IMC), Global Cosmics, and Beamline for Schools (BL4S). In addition, a training program at Fermilab and other U.S. labs serves Italian/European graduate students in physics and engineering. IMC is the largest of these programs, reaching over 10,000 high school students each year with authentic data from the collider and neutrino experiments as well as proton therapy simulations. Global Cosmics is the umbrella for a large number of activities that use cosmic ray detectors and data they produce. BL4S is a competition sponsored by CERN for high school students to propose experiments to be run in a CERN (or, recently, DESY) fixed-target beamline; the winners of the competition are able to travel to the accelerator and, with professional support, conduct their experiment.

### 4.1. International Masterclasses

International Masterclasses are, most commonly, single-day events in which students and teachers are invited to a university or laboratory to spend an intense day learning about particle physics. The highlight is 1-2 hours in which students analyze data from a contemporary particle physics experiment with mentoring from physicists. The goal is to enable the students to be "particle physicists for a day". At the end of the day, the students and teachers participate in a videoconference moderated by physicists originally at CERN; Fermilab joined in moderating IMC videoconferences in 2008, TRIUMF, KEK, or GSI have joined this effort since then. [22]

IMC started with LEP data in 2005, using OPAL and DELPHI event displays pioneered for student use in the UK and Sweden. It gained immediate popularity in Europe. The U.S. joined the effort in 2006 with two masterclass institutes at Brookhaven National Laboratory and Florida State University; U.S. masterclasses have since grown to about 20-30 U.S. institutes each year. In 2010 IMC began the switch to LHC data: the CMS masterclass measurement was built and is maintained by U.S. physicists and QuarkNet. In 2018, neutrino masterclasses based on the Fermilab MINERvA experiment were added to IMC and more neutrino masterclass measurements are expected in the coming 1-2 years. Since

2012, IMC has been co-coordinated by Uta Bilow at Technische Universität Dresden under the aegis of CERN and IPPOG and Kenneth Cecire at the University of Notre Dame as a staff member of QuarkNet. [23]

Three auxiliary pieces of IMC were developed in the U.S. by QuarkNet. The first is World Wide Data Day (W2D2), a simpler masterclass-type event which can be done by students in high school classrooms with coaching by physics teachers and culminates in videoconferences with physicists. All videoconferences take place on the one eponymous day each autumn but teachers and students can analyze data well before that if they choose. The entire time commitment is about 2 hours. Another simplified version, the Big Analysis of Muons (BAM), takes place in the late spring. It started in 2020 as a response to the cancellation of most masterclasses due to COVID-19 to give teachers a chance to engage their students in a masterclass while teaching remotely. BAM is popular in some parts of the world where traditional masterclasses have not taken firm root, perhaps due to lack of access to nearby particle physics institutions. Both W2D2 and BAM use ATLAS and CMS data. [24]

The third important auxiliary to IMC is Masterclasses for the International Day of Women and Girls in Science, focused on providing role models and discussions to encourage young women in physics. These are almost all done in Europe but it should be possible to have some in the U.S. as well in the future. [25]

A great advantage of strong U.S. participation in IMC is its international nature. Students and teachers gain access to data from leading experiments at CERN and collaborate online with their peers around the world who analyze the same data. At the same time, the U.S. is a good collaborator by adding organizational infrastructure and, increasingly, data from U.S. experiments such as the MINERvA neutrino experiment. IMC has redounded to the public benefit of U.S. particle physics due to the fairly large number of foreign masterclass institutes that connect with Fermilab for videoconferences and cooperate with the U.S. and QuarkNet for information, orientations, and assistance.

More particle physics experimental groups in the United States should offer support to International Masterclasses in the United States by hosting masterclasses; more individual particle physicists should offer expertise as tutors in masterclasses as well as moderators for masterclass videoconferences. The particle physics community should encourage this

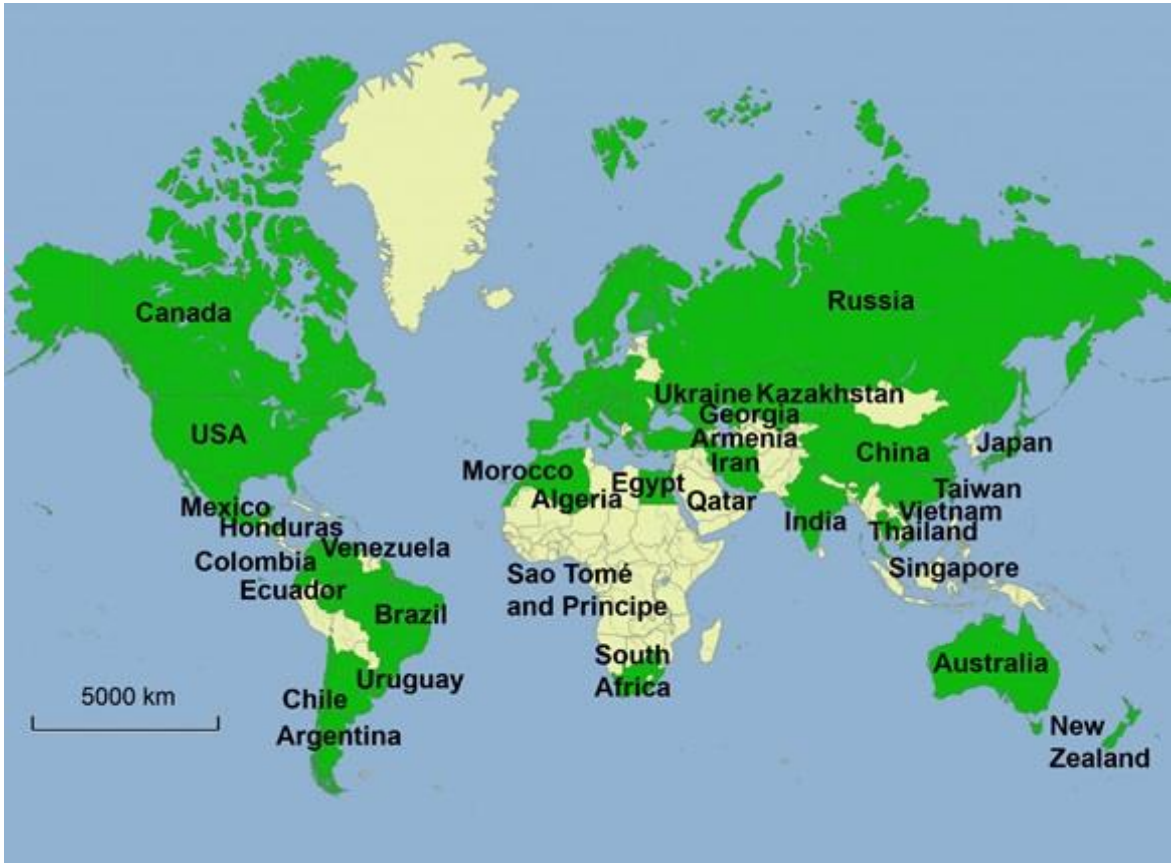

Figure 3: Distribution of countries worldwide participating in International Masterclasses. (Image by Uta Bilow.)

**Finding:**

F3. International Masterclasses are an excellent model of international physics collaboration for high school students and teachers. Increased promotion of and participation in IMC by particle physics experimental groups in the U.S. would benefit the particle physics community.

## 4.2. Global Cosmics

Cosmic ray devices and their associated educational programs are an excellent means of reaching and connecting diverse classrooms across the country. Schools with financial means are able to install and operate the devices, while those with limited resources can be granted access to the data, sharing and comparing results on an equal basis with the others. QuarkNet has made a significant effort to develop affordable systems and to help to install them in classrooms across the U.S. They have also created and supported networks connecting the classrooms. This program, along with others set up in the U.S. and around the globe, provide students with affordable experiences with real particle physics detectors and data without requiring the presence of active research centers or institutes.

The Global Cosmics portal [26] is hosted by DESY with the goal of unifying various worldwide classroom programs. The goal of the portal is to unify access to the detectors and the data, allowing students from classrooms with detectors or without detectors to learn and share experiences. The portal is relatively new and will be coming online with a new, more accessible interface in 2022, allowing for easier access to diverse worldwide audiences. The QuarkNet Cosmic network is one of the key members of the IPPOG Global Cosmics portal.

QuarkNet should continue to develop and distribute affordable, maintainable cosmic ray devices for classroom use. It should augment its efforts to connect classrooms across the country, with an emphasis on reaching lower income and underrepresented communities. This effort should be integrated with the Global Cosmics portable to create an affordable means for students in the USA to work alongside their counterparts overseas. Programs connecting students through events, such as eclipses, should be continued and broadly advertised throughout the country.

**Finding**:

F4. Development and distribution of affordable, maintainable cosmic ray devices for classroom use internationally and in the U.S. have increased collaboration in cosmic ray studies. This should be expanded.

### 4.3. Beamline for Schools

Beamline for Schools (BL4S) is an international competition for high school age students held each year by CERN. Teams of students from around the world propose experiments for a beamline from the Proton Synchrotron. CERN physicists review these proposals and pick 1-2 winners each year, who are then brought to CERN, expenses paid, to work with physicists to carry out their winning experiment. [27] Since the program began in 2014, eleven teams have won the competition, including one from the United States. [28] Several U.S. teams have also been recognized for outstanding proposals. Two of these were able to carry out their experiments at the Fermilab Test Beam Facility. U.S. teachers and students who participated were very positive about their experiences whether or not they received any award: the process of working together to research, design, and present an experiment made for a great opportunity. [29]

**Finding**:

F5. The U.S. particle physics community can encourage participation in Beamline for Schools by arranging for a number of U.S. teams that do not win to carry out their experiments in U.S. facilities. They can expand these benefits to more students by creating an international research competition to complement Beamline for Schools at one or more U.S. laboratories; this will also highlight U.S. research to students and the public.

**Recommendation**:

R3. Continue to support efforts by U.S. teams in Beamline for Schools and provide analogous opportunities within the U.S.

### 4.4. STEAM Initiatives

STEAM - Science, Technology, Engineering, Arts, and Mathematics - is a way to bring the arts into motivating and giving new insights to students in more traditional STEM fields on one hand and of opening doors to students who are on a very different path to the inspiration inherent in scientific methods and discovery. [30]

Through her STEMarts Lab program [31], artist Agnes Chavez is a leader in STEAM with high school students in New Mexico, many of them indigenous Americans. She has written:

*The field of particle physics and technological advancements are radically transforming our understanding of the universe and pushing the boundaries of how we live, work and communicate. At the same time we are faced with planetary crises such as climate change, pandemic diseases, and massive social inequities in access across borders. Our youth and our communities from diverse socio-economic and cultural populations need to be informed and empowered to respond to these 21st century challenges.*

*Through our STEAM-based interdisciplinary practice, we have seen that the development of scientific literacy, coupled with artistic and humanistic literacy, is key to prepare our future youth leaders for this new world. Students learn to use art and humanities as a 'way of knowing' and understand how fundamental research in physics can expand their minds and provide the tools they need to reimagine humanity and build a better future.*

*We found that students emerged from this experience with a deeper appreciation of the potential for art / science / humanities inter-relationship, a greater sense of purpose in developing opportunities for applying their new knowledge, and renewed hope in a future that they have the power to create.*

Chavez goes on to describe a case study in STEAM and international engagement:

The COVID pandemic has thrown our educational institutions into upheaval, but at the same time through the proliferation of virtual and augmented realities, it has provided an opportunity for more accessible and integrated pedagogical systems. During the lockdown, STEMarts Lab added a virtual international exchange program to address the isolation and depression rising in the schools. Students in underserved rural communities of New Mexico collaborated with students in Portugal through a 7-week interdisciplinary workshop that explored the universe through the lens of art, particle physics, astrophysics, native science and philosophy. Through conversations with experts, they experienced why it matters to gain knowledge in fundamental research in physics, art and philosophy. Their contributions became part of a final 'Mixed reality' installation that is presented at art and science festivals around the world to inspire scientific literacy.

It should be noted that, previous to COVID, Chavez and high school students from the Taos area participated in a very successful ATLAS masterclass working with QuarkNet teacher Dr. Michael Wadness of Medford High School, Massachusetts. Since then, she has worked on STEAM programs with QuarkNet Staff Member Shane Wood.

### 4.5. International Schools

This section covers two kinds of International School. The first, which we will call International Graduate (IG) Schools, are extended workshops presented in various venues around the world. Well-known examples include CERN-Fermilab Summer Schools [32] and the African School of Physics. [33] The second are schools located around the world for high school students from outside the host country, although these schools also accept local students, often in large numbers. We will refer to these as International Secondary (IS) Schools, as we will focus on the high school level for such schools. These two very different types of school have very different relationships to international connections in physics education.

IMC has been successful in expanding awareness of masterclasses, especially to younger physicists, through IG Schools and international meetings. This began as an adjunct activity to the November 2019 IPPOG meeting and the pattern has held since then. Physicists are invited to two 45-minute sessions in which various masterclass measurements are explained and sampled. Each registrant can therefore sample two measurements. This was since repeated in another IPPOG meeting as well as three additional venues, all virtual, including the APS April 2021 Meeting. [34]

The other IG School involvement is deeper. QuarkNet has, along with international partners, been involved in the education and outreach extension of the African School of Fundamental Physics and Applications (ASP). QuarkNet and partners participated in high school teacher and learner programs in ASP 2016 in Kigali, ASP 2018 in Windhoek, and ASP 2020 virtual meetings (originally scheduled for Marrakech). In the first two, QuarkNet and partners held physics education workshops for teachers and visited schools to introduce learners to particle physics. In addition, they brought cosmic ray detectors for the graduate students to experiment with and to demonstrate to high school learners. They also ran optional masterclasses for the graduate students to experience. In the virtual version, held online in 2020 and 2021, QuarkNet presented physics education lectures for the graduate students and a virtual workshop for African teachers.

QuarkNet collaborates with a loose network of IS Schools, both within and outside of the U.S., to introduce and share their programs. This partnership has extended the reach and audience of the programs to a global scale. International schools are often seen as havens of innovation and therefore are natural extensions of the goals of the outreach programs. IS Schools in Shanghai, Qingdao, Fukuoka, Singapore, Amman, and Riyadh have all participated in various ways in masterclasses, affecting both American teachers and students at those schools and participants in U.S. masterclasses that partnered with them in videoconferences. The school in Shanghai is involved to the extent that the physics teacher there is a QuarkNet fellow who also maintains a cosmic ray detector with his students and contributes to QuarkNet activities.

**Findings:**

F6. The masterclass introductory sessions pioneered in international meetings and schools build interest in not only masterclasses but particle physics education and outreach when carried to graduate student and young physicist settings in the U.S. and abroad.

F7. The QuarkNet relationship with the African School of Fundamental Physics and Applications should become a permanent feature; QuarkNet should also seek collaboration with other relevant initiatives from developing nations or regions.

## 5. STUDENT OPPORTUNITIES ACROSS BORDERS

### 5.1. CERN REU Summer Student Program

Every (non-pandemic) year since 1962, CERN has run a Summer Student program [35], offering an opportunity for undergraduate students in its member states to come to Geneva to work with top researchers in physics, engineering and computing for 2-3 months between semesters. The program has been tremendously successful and has grown in size every summer. Typically, around 3000 students apply for entry to the program, which accepts around 280.

In 1998, funding was secured through the NSF Research Experience for Undergraduates (REU) program to include U.S. students. This program, administered by the University of Michigan, brings 15 students selected from across the U.S. to CERN for 9 weeks every summer. The program has been renewed every year since. Its success is evident by its popularity (roughly 300 students apply each year for 15 places) and by the success rate of students who have taken part in the program, as measured through surveys, fraction of students attending graduate studies, and the success of those students in their career path.

Although the number of students permitted each summer is limited by the constraints of the CERN non-member state program and by the logistics of housing and mentor availability, additional students have participated in semester research programs (see below), which are organized in a similar manner, but with more in-depth, longer-term projects.

### 5.2. CERN Undergraduate Semester Research Program

Since 2013, the University of Michigan has hosted a semester-long research program for undergraduate students at CERN [36]. The students are selected from a diverse mix of small and large universities across the U.S. and are embedded as CERN Users in active research programs on experiments at the laboratory. The program is modeled on the highly successful NSF-funded Research Experience for Undergraduates (REU) CERN Summer Student program (see above), but is typically around 100 days long, providing the students with a much richer research experience.

Motivation for the program came not only to address overflow from the summer program, but also to serve students in need of semester research credits, to reach out to more diverse communities, to reach out to typically lower-income students constrained by summer work obligations, and to explore the advantages of a longer, more in-depth experience. For the program, students are registered as official CERN Users and are embedded in international research teams exactly as any other scientist at the lab.

The mentors are selected based on their leadership skills, as well as their ability to educate and inspire the students. Projects cover a wide range of activities from detector R&D to software development, trigger design, physics analysis and theoretical methodology, touching nearly all aspects of the research program at CERN. The students live in apartment facilities in nearby St. Genis Pouilly, and enjoy periodic excursions to cultural centers located around Europe.

The number of students participating has varied over the years, depending on the funding source. Funding for the first 5 years came from the Richard Lounsbery Foundation, supporting 5-6 students each semester. Although feedback was exceptionally positive, the foundation limits itself to supporting start-ups, and could not continue. Since Fall 2018, the University of Michigan Department of Physics has supported 3 students from the University for one semester each year. In addition, a grant has been secured by the State Department through the United States Mission to the International Organizations in Geneva to support another 3 students each year. These students are selected nationwide and are either women or from communities that are traditionally underserved in the field.

### 5.3. DOE-INFN Summer Students Exchange Program

The U.S. Department of Energy (DoE) and the National Institute for Nuclear Physics - Italy (INFN) run a joint Summer Students Exchange Program, dedicated to the exchange of U.S. and Italian students in science and engineering. [37] U.S. students who are willing to join a INFN research team operating in Italy can spend 2 months at any of the INFN sites. Candidates must be enrolled as students at a U.S. university and they must have begun, at the time of application, at least the third year of a U.S. university curriculum in physics, engineering or computing science.

There are 11 grants of Euro 5000 each for covering travel and living expenses.

### 5.4. Training in the U.S. of Italian/European Graduate Students in Physics and Engineering

Since 1984, INFN and University of Pisa scientists performing experiments at Fermilab have been running a two-month summer training program for Italian students at the lab. In addition, based on the strong interest that a number of former Italian summer students showed to come back to the lab and continue their research, E. Barzi started in 1998 what later became an official thesis program for an Italian Laurea Magistrale. A number of these students are now permanent Fermilab employees. This shows how the laboratory had a significant return from this program. [38-43].

The symmetric "DOE-INFN Summer Students Exchange Program" is described above. In 1984 the program involved only a few physics students from the University of Pisa, but it was later extended to other INFN groups and to engineering students. Since 2004 the program has been supported in part by DOE in the frame of an exchange agreement with INFN and has been run in part by the Cultural Association of Italians at Fermilab (CAIF, [44, 45]). In 2007 the Sant'Anna School of Advanced Studies (Pisa) established an agreement with Fermilab to share the cost of four engineering students each year. In the 36 years of its history, the program has hosted at Fermilab approximately 550 Italian students from more than 20 Italian universities and from some non-Italian universities. Several of these students returned to the U.S. for their PhD and are now employed as permanent staff at Fermilab and other labs. In addition, in the years 2010-2020, with the support of the National Institute of Space Physics (INAF), the Italian Space Agency (ASI), and CAIF, 30 students were hosted in other U.S. laboratories and universities. The steady increase of the trainees with time is shown in Fig. 4.

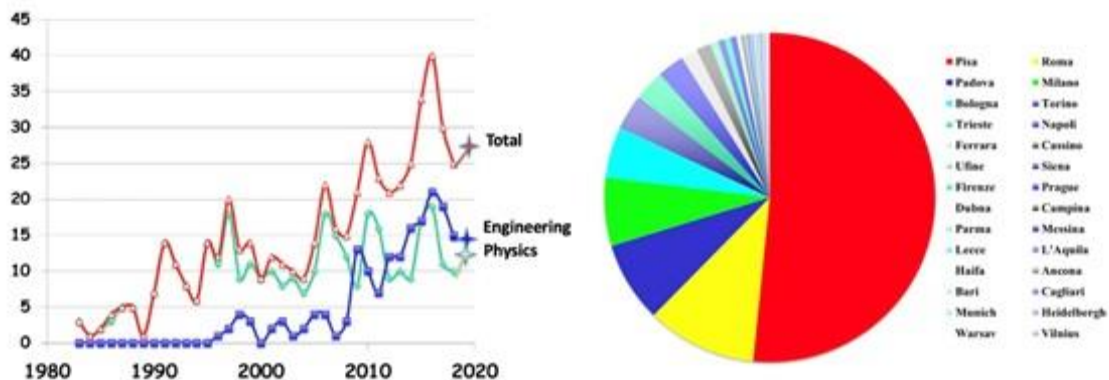

Figure 4: (left) Distribution of the number of students selected in the years 1984-2019 (the program of the year 2020 was canceled due to COVID-19): the blue curve represents the distribution of the engineering students, the green curve represents the distribution of the physics students and the red curve represents the total distribution. One notes an increased rate of engineering students in recent years, who are now as many as the physics students; (right) Frequency of accepted students as a function of the university of origin, including two non-Italian universities.

Trainings at Fermilab span over the entire range of science and technology programs of the laboratory. These include local neutrino and rare particle physics experiments, the CMS experiment at CERN, astrophysical searches and analysis of astrophysical data, R&D on accelerator technologies, particle physics theory, design and construction of particle detectors and accelerator components, advanced computing and advanced data handling, superconducting R&D, and theory of accelerators. Reading the final reports in the lab web site is very instructive on the importance of student contributions.

The success of these programs is due to the high quality of the recruited students, to their assignment of appropriate programs, and importantly, to continuous monitoring of their work quality. The program team carefully follows the students. After one month, they present an oral report on the progress of their project, which they follow with a final oral report at the end of the period. Before they depart, the students present written papers describing the results of their work and to comment on what they have learned during the stage. After approval, the final written reports are saved as internal Fermilab scientific publications. As such they become part of the student`s curricula. Upon their return to Italy, the students pass an exam with University of Pisa professors and earn credits for their university course. Most universities recognize these credits.

In 2019, two students were hired at BNL to perform a similar two-month long summer training stage on particle physics science and technology as at Fermilab. Two BNL research groups interviewed students from the pool of selected ones, and offered to two of them a training period on ongoing projects of BNL, with conditions very similar to those offered at Fermilab. They completed their programs very successfully. Hopefully this important extension at BNL of students training in physical science and technology in the U.S. will be continued in the future to attract international partners. In order to do this, Fermilab should expand its FWP proposal to include other DOE labs with interests in HEP that are interested in joining the program.

**Finding:**

F8. The particle physics community should make every effort to seek out participation in the CERN, Fermilab, and other international summer and semester programs by U.S. applicants from underserved communities, including minorities and those of lower income families. Additional support could come from dedicated NSF, APS and State Department. Programs. Expansion of these programs would benefit all students and the wider community.

**Recommendation:**

R3. U.S. institutions should expand and more fully support international student research and learning opportunities for both students coming to the U.S. and U.S. students going abroad.

## 6. INTERNATIONAL SCIENCE DIPLOMACY

This paper takes no position on questions of diplomacy but assumes that understanding and mutual respect between people of the United States and other countries is a general good. Such understanding and respect can be enhanced by positive interactions between U.S. and other scientists, science teachers, and students at various levels. Sharing of resources and knowledge between the U.S. and developing countries can be particularly helpful in this regard. This includes not only bringing what we have learned to our colleagues in the developing world but also an ability to listen to the actual needs of those colleagues and openness to what they can help us to learn from their perspectives.

In particle physics, there are many opportunities for researchers to get to know and appreciate each other. This also occurs at the graduate level with IG Schools, as explained above. These form a natural basis for international science diplomacy. As seen in much of this paper, extension of such opportunities to teachers and undergraduate or high school students can and often does follow this paradigm. Well-designed programs of outreach to other countries, especially those in the developing world, can have a great impact because they appeal to a wider spectrum of citizens and can deliver impactful assistance in physics education.

International Masterclasses and, where practical, cosmic ray studies programs can form the backbone of such efforts. The spread of masterclasses in Latin America has been a testament to this. In addition, QuarkNet Cosmic ray detectors and the network that goes with them are sited and available to physicists, teachers, and students in Chile, Ecuador, and Mexico. Current work with the African School of Fundamental Physics and Applications lays the groundwork for similar growth in Africa.

These and other ongoing programs can be expanded for purposes of international diplomacy if the interest and opportunity arise.

**Finding:**

F9. Education and outreach programs of mutual benefit help form strong collaborations by theoretical particle physicists or institutions in the U.S. with developing nations and should be encouraged.

## 7. CONCLUSIONS

International engagement in particle physics education and outreach is a natural outgrowth of the international nature of particle physics itself. This has multiple benefits for students and teachers in terms of their understanding of physics practice and their appreciation of the whole educational enterprise. Consideration of CERN and its international collaborations, for example, not only enhances understanding of physics and engineering but also of foreign languages, history, and culture.

International engagement adds richness to physics education as well as helps educators find best practices through collaboration with each other across borders and oceans. Teachers discover what experiences they have in common and what is uniquely their own. The opportunity arises to assist those in developing countries and learn from them in turn. Students discover that their peers on one hand and scientific experts on the other come from every possible cultural background and that they have much to share with each other.

In a nutshell, we do particle physics education and outreach much better because of the international connections we have and those we seek. The U.S., as a leading country and a multi-ethnic republic, benefits greatly and offers much from this. The U.S. particle physics community should encourage and build on its leadership and example in this area.

**Finding:**

F10. An international particle physics summer school for teachers in the U.S. with a strong but accessible content orientation, along with emphasis on physics education reform, would be a unique contribution to education while complementing other international efforts.

**Recommendation:**

R5. Establish, with international partners, an International Particle Physics Summer School for Teachers.

## Acknowledgments

The authors wish to thank Randal Ruchti of the University of Notre Dame and Sudhir Malik of the University of Puerto Rico Mayagüez for their guidance in and assistance in preparing this Contributed Paper. They would also like to thank the many physicists, educational specialists, teachers, and students at various levels for showing the international particle physics education by the example of their work.